\documentstyle[preprint,prd,aps]{revtex}

\def\gsim{\begin{array}{c} > \\ \sim \end{array}}

\begin{document}
\draft
\title{\large \bf Energy-momentum uncertainties as possible origin of
threshold anomalies in UHECR and TeV-$\gamma$ events}
\author{\bf Y. Jack Ng$^{(a)}$\thanks{Corresponding author. E-mail:
yjng@physics.unc.edu},
D.-S. Lee$^{(b)}$, M. C. Oh$^{(a)}$, and H. van Dam$^{(a)}$}
\address{(a) Institute of Field Physics, Department of Physics
and Astronomy,\\
University of North Carolina, Chapel Hill, NC 27599-3255\\}
\address{(b) Department of Physics, National Dong Hwa University,\\
Shoufeng, Hualien 974, Taiwan, Republic of China\\}
\maketitle
\bigskip

\begin{abstract}

A threshold anomaly refers to a theoretically expected energy threshold
that is not observed experimentally. 
Here we offer an explanation of the threshold
anomalies encountered in the ultra-high energy cosmic ray events and
the TeV-$\gamma$ events, by arguing that energy-momentum uncertainties
due to quantum gravity, too small to be detected in low-energy regime, can
affect particle kinematics so as to raise or even eliminate the 
energy thresholds. A possible
modification of the energy-momentum dispersion relation, giving rise to
time-of-flight differences between photons of different energies from gamma
ray bursts, is also discussed.

\bigskip


\end{abstract}

\newpage
\section{Introduction}

The observation of ultra-high energy cosmic rays\cite{crob} with energy
exceeding the
Greisen-Zatsepin-Kuz'min cutoff\cite{GZK} at $\sim 5 \times 10^{19}$ eV has
presented
the physics and astrophysics community quite a conundrum. The GZK cutoff is
based on pion photo-production by inelastic collisions of cosmic-ray
nucleons with the cosmic microwave background
\begin{equation}
p + \gamma (CMB) \longrightarrow p + \pi.
\label{pgam}
\end{equation} 
(Actually, the dominant contribution to the GZK cutoff comes from the
$\Delta(1232)$-resonance. But the difference between $m_\Delta$ and $m_p +
m_\pi$ would modify our results below only slightly. Moreover, if the
$\Delta$ formation is not possible, a weakened version of the GZK cutoff
may result from non-resonant pion photo-production. Also we should add
that the exact composition of the cosmic rays is not known. But even if
they are heavy nuclei like Fe rather than nucleons, they would still be photo-disintegrated, and the GZK cutoff remains
more or less intact.) In the CMB frame, such a collision requires a
threshold energy of the cosmic ray proton given by
\begin{equation}
E_{th} = \frac{(m_p + m_{\pi})^2 - m_p^2}{4 \omega} \simeq 5 \times 10^{19} \hspace{0.1in} eV,
\label{crth}
\end{equation}
for an average CMB photon energy $\omega \sim 1.4 \times 10^{-3}$ eV (and
$m_p \simeq 9.4 \times 10^8$ eV for the proton mass, $m_{\pi}
\simeq 1.4 \times 10^8$ eV for the pion mass). For
protons above this energy, the pion photo-production from CMB will dominate
beyond the mean free path. And for pion photo-production cross-section of
$\sim 200 \mu b$ and density $\sim 550$ photons/cc for the CMB, this mean
free path is
of order 1 Mpc, much smaller than the intergalactic distances. This would
imply that the protons would need to originate within our galaxy. But the
known maximal galactic magnetic fields are too weak to accelerate the
protons to such ultra-high energies. Furthermore, such high energy $\sim
10^{20}$ eV protons are hardly deflected by the interstellar magnetic
fields and hence should have a direction identifiable with some source.
But the observed UHECR events are oriented along the extragalactic plane
and have no known correlation with any identifiable sources. Thus we are
forced to conclude that nature has found a way to evade the GZK cutoff, an
energy threshold that, as we have just seen, is well established
theoretically. This phenomenon of a theoretically expected energy
threshold not observed experimentally has come to be known as a threshold
anomaly.

There has not been a lack of attempts\cite{Oli} to explain these
extraordinary
cosmic rays. They include protons originated from nearby topological 
defects/monopolium\cite{topde}, magnetic monopoles \cite{mono}, and
solutions like the 
decay of massive supersymmetric hadrons\cite{S0}. Other explanations
include 
"Z-boson bursts"\cite{Zbur} and decay products of hypothetical super-heavy
relic 
particles\cite{Relic}. Exotic origins have also been suggested, such as: 
gamma-ray bursts\cite{gambur}, spinning
supermassive 
black holes associated with presently inactive quasar
remnants\cite{spinbh}, 
and colliding galaxy systems\cite{colga}.

The recent observation of 20 TeV $\gamma$-rays\cite{HEGRA} from Mk 501 is
also puzzling.\cite{PM} Theoretically
 such events are not expected since a high energy
photon propagating in the intergalactic medium can suffer inelastic impacts
with photons in the Infra-Red background resulting in the production
of
an electron-positron pair
\begin{equation}
\gamma + \gamma (IR) \longrightarrow e^{+} + e^{-}.
\label{gaga}
\end{equation}
For such a collision, the threshold energy of the high energy photon is
given by 
\begin{equation}
E_{th} = \frac{m_{e}^2}{\omega} \simeq 10 \hspace{0.1in} TeV,
\label{gath}
\end{equation}
for an average photon energy of $\omega \sim 0.025$ eV in the IR background (and 
$m_e \simeq 0.5 \times 10^6$ eV for the electron mass). Thus $\gamma$-rays 
above 10 TeV lose energy drastically during their propagation from their 
source to the Earth. It is very unlikely that they can survive their long 
trip from distant Mk 501 with any significant flux. Here then
is another threshold anomaly. Compared to the UHECR events, the
TeV-$\gamma$ events have elicited only a few explanations, such as: there may be a possible upturn in the intrinsic spectrum emitted by Mk 501; 
the distance to Mk 501 or the 
background IR intensity may have been overestimated; and multiple 
TeV-$\gamma$ emitted coherently by Mk 501 may have been mistaken to be a 
single photon event with higher energy\cite{gaexp}.

There is one solution to the UHECR paradox and recently used also to
explain the TeV-$\gamma$ events that deserves special mention. Numerous
authors\cite{LID,colga,ACP} have suggested that these events are a signal
of violation of
ordinary Lorentz invariance at the energies in question. These violations
are too small to have been detected at the available accelerator energies.
But at the highest observed energy region they can affect particle
kinematics so as to suppress or even forbid the inelastic collisions (Eq.
(\ref{pgam}) and (\ref{gaga})), thereby evading the two cutoffs.

In this paper we will adopt a proposal\cite{ng}, which bears some
similarity
to the one just mentioned, to solve the UHECR and TeV-$\gamma$ puzzles. It
is based on the observation that, due to quantum gravitational effect,
energy and momentum, like distances and time intervals, cannot be measured
with infinite accuracies. The energy-momentum uncertainties of
the form
(with positive $a$) 
\begin{equation} 
\delta E \gsim E \left(\frac{E}{E_P}\right)^{a}, \hspace{0.5in} \delta p \gsim p \left(\frac{p}{m_P c}\right)^{a}, 
\label{emun} 
\end{equation} 
a natural consequence of quantum gravitational effects,\cite{nvD1} 
upon inserted into the energy-momentum conservation
equations or the energy-momentum dispersion relation, can mimic the effects
of violation of ordinary Lorentz invariance in a particular way. (Here
$E_P$ denotes the Planck energy, $m_P$ denotes the Planck mass, and we have restored the factor of $c$.)  They
can be interpreted as the \emph{physical} origin of the threshold
anomalies. We have little to say about the origins of UHECR and
TeV-$\gamma$ per se. We simply want to point out that there is a
\emph{natural} mechanism that can potentially raise or even eliminate the
two energy thresholds under consideration.

The outline of the paper is as follows: In the next Section, we review the
argument used by two of us (Ng and van Dam)\cite{nvD1} years ago leading to
energy-momentum uncertainties of the form given by Eq. (\ref{emun}). 
In Section III, we use the
energy-momentum uncertainties to explain the threshold anomalies encountered 
in the UHECR and TeV-$\gamma$ events. We also give
another plausible interpretation of energy-momentum uncertainties and apply
it to future time-of-flight measurements of photons of different energies
from gamma ray bursts. The concluding section is devoted for discussion.

\section{Energy-momentum uncertainties}

Just as there are uncertainties in distance and time interval measurements,
there are uncertainties in energy-momentum measurements.  Both types 
of uncertainties\cite{nvD1}
come from the same source, viz., quantum fluctuations of space-time
metrics\cite{Ford} giving rise to space-time foam. We will consider 
two leading 
models of space-time foam. In the first model, the
fluctuations of the metric are given by\cite{MTW}
\begin{equation}
\delta g_{\mu \nu} \gsim \frac{l_P}{l},
\label{mtwdg}
\end{equation}
for a measurement in a space-time
region of volume $l^4$. Here $l_P \equiv (\hbar G/c^3)^{1/2}$ is the Planck length.
Since $\delta l^2 = l^2 \delta g$, this translates into an
uncertainty in distance measurements given by $\delta l \gsim l_P$. We can
calculate the minimum uncertainty in momentum for a particle with momentum
$p$ by regarding $\delta p$ as the uncertainty of the momentum operator $p =
-i \hbar \partial / \partial x$, associated with $\delta x = l_P$. For any
function $f(x)$, $(\delta p) f$ is given by
\begin{equation}
(\delta p) f = \frac {\hbar}{i} \left(\delta x \frac{\partial^2 f}{\partial x^2} 
+ \frac{\partial f}{\partial x} \frac{\partial \delta x}{\partial x}\right).
\label{dpf}
\end{equation}
Taking the function $f(x)$ to be a momentum eigenstate $f =
exp(ipx/\hbar)$, we get
\begin{equation}
(\delta p) e^{ipx/\hbar} = i \frac{p^2 l_P}{\hbar} e^{ipx/\hbar}.
\label{dpe}
\end{equation}
This yields
\begin{equation}
|\delta p| \sim p \left(\frac{p}{m_P c}\right),
\label{dpm}
\end{equation}
where $m_P \equiv (\hbar c/G)^{1/2}$ is the Planck mass. 

An alternative derivation of Eq. (\ref{dpm}) goes as follows: Imagine
sending
a particle of momentum $p$ to probe a structure of spatial extent $l$ so that 
\begin{equation}
p \sim \frac{\hbar}{l}.
\label{pun}
\end{equation}
Consider the coupling of the metric to the energy-momentum tensor of the
particle,
\begin{equation}
(g_{\mu \nu} + \delta g_{\mu \nu}) t^{\mu \nu} = g_{\mu \nu} (t^{\mu \nu} 
+ \delta t^{\mu \nu}),
\label{stt}
\end{equation}
where we have noted that the uncertainty in $g_{\mu \nu}$ can be translated into
an uncertainty in $t_{\mu \nu}$. Eqs. (\ref{mtwdg}) and (\ref{stt}) can now be used to
give
\begin{equation}
\delta p \gsim p \left(\frac{l_P}{l}\right),
\label{dp}
\end{equation}
which, with the aid of Eq. (\ref{pun}), yields Eq. (\ref{dpm}). We can also
mention that the momentum uncertainty is actually fixed by dimensional analysis,
once the uncertainty in the metric is given by Eq.
(\ref{mtwdg}). The corresponding statement for energy uncertainties is 
\begin{equation}
\delta E \sim E \left(\frac{E}{E_P}\right).
\label{de}
\end{equation}

Next let us consider the second space-time foam model\cite{nvD1,KS} (which we
actually favor over the first model for reasons we have given 
in Ref. \cite{ng,nvD2}, including its natural connection to the holographic 
principle and black hole physics). It is given by
\begin{equation}
\delta g_{\mu \nu} \gsim \left(\frac{l_P}{l}\right)^{2/3},
\label{nvddg}
\end{equation}
corresponding to $\delta l \gsim (l l_P^2)^{1/3}$. Repeating the above
procedure we get
\begin{equation}
\delta E \gsim E \left(\frac{E}{E_P}\right)^{2/3}, \hspace{0.5in} \delta p 
\gsim p \left(\frac{p}{m_P c}\right)^{2/3}.
\label{dedp}
\end{equation} 

Note that, for both space-time foam models, the
energy-momentum uncertainties are negligible except when we consider
processes involving very energetic particles. We should also mention that
we have not found the proper (presumably nonlinear) transformations of 
the energy-momentum uncertainties between different reference frames.
Therefore we will apply the results only in the frame in which we do the
observations. In the following, we will write the energy-momentum
uncertainties in the form given by Eq. (\ref{emun}) with
$a = 1, 2/3$ for the space-time foam models given by Eq. (\ref{mtwdg}) and
Eq. (\ref{nvddg}) respectively.

\section{Solving the threshold anomalies}

Now that we know the energy-momentum uncertainty expressions, we have to
figure out how
and where we should apply them. It all comes down to the question of correctly
interpreting the physics. Relevant to the discussion of the UHECR events
and the TeV-$\gamma$ events is the scattering process in which an energetic 
particle
of energy $E_1$ and momentum $\mathbf{p}_1$ collides head-on with a soft
photon of
energy $\omega$ in the production of two energetic particles with
energy
$E_2$, $E_3$ and momentum $\mathbf{p}_2$, $\mathbf{p}_3$. Henceforth let us adopt 
$c = 1$. At threshold,
(ordinary) energy-momentum conservation demands
\begin{equation}
E_1 + \omega = E_2 + E_3, \hspace{0.5in} p_1 - \omega = p_2 + p_3,
\label{oemc}
\end{equation}
and the (ordinary) energy-momentum dispersion relation takes the form
\begin{equation}
E_i = ( p_i^2 + m_i^2)^{1/2},
\label{oemd}
\end{equation}
where $i = 1,2,3$ refers to the particle with energy $E_i$,
momentum $p_i$, and mass $m_i$.  For the UHECR and TeV-$\gamma$ events, 
these two equations yield the threshold energies given by Eqs. (\ref{crth}) 
and (\ref{gath}) respectively.  But for the problem of threshold anomalies 
at hand, we believe Eqs. (\ref{oemc}) and (\ref{oemd}) can receive crucial
modifications from energy-momentum uncertainties.\cite{ng}
Let us, therefore, consider (separately) modifying 
(i) the conservation expressions and (ii) the dispersion relation.  (The 
suggestion that the dispersion relation may be modified by quantum 
gravity first appeared in Ref.\cite{IJMP}.)

(i) Modifying energy-momentum conservation relations

While the energy-momentum dispersion relation is the conventional one given
by Eq. (\ref{oemd}), 
\begin{equation}
E_i \simeq p_i + \frac{m_i^2}{2 p_i},
\label{onemd}
\end{equation}
where we have used the fact that $p_i$ is very large compared to $m_i$,
the energy-momentum conservation is modified to read
\begin{equation}
E_1 + \delta E_1 + \omega = E_2 + \delta E_2 +E_3 + \delta E_3,
\label{mec}
\end{equation}
and
\begin{equation}
p_1 + \delta p_1 - \omega = p_2 + \delta p_2 + p_3 + \delta p_3.
\label{mmc}
\end{equation}
Thus, in this scheme, energy-momentum is conserved up to 
energy-momentum uncertainties, while the dispersion relation is still dictated by Lorentz invariance. 
We have omitted $\delta \omega$, the contribution coming from the
uncertainty of $\omega$ because $\omega$ is small.
Substituting Eq. (\ref{onemd}) into Eq. (\ref{mec}) and making use of Eq.
(\ref{mmc}),
we can rewrite Eq. (\ref{mec}) as
\begin{equation}
4 \omega \simeq \frac{m_2^2}{p_2} + \frac{m_3^2}{p_3} - \frac{m_1^2}{p_1} 
+\varepsilon \frac{1}{E_P^a} (p_1^{1 + a} - p_2^{1 + a} - p_3^{1 + a}).
\label{lmcr}
\end{equation}
Here we have used Eq. (\ref{emun}) and the fact that $E_i \simeq p_i$ for
energetic particles to put
\begin{equation}
\delta p_i - \delta E_i \simeq \varepsilon \frac{p_i^{1 + a}}{2 E_P^a},
\label{epsd}
\end{equation}
thereby defining the parameter $\varepsilon$.  We do not know how to
calculate $\varepsilon$; but since $\delta E_i
\simeq \delta p_i$ for high energy, we expect that it 
can be fairly \emph{small} compared to one. 

The solution to Eq. (\ref{lmcr}) for the threshold energy 
$E_{th} \simeq p_1$ of the
incoming energetic particle can be easily worked out for
the case of TeV-$\gamma$ for which $m_1 = 0$, $m_2 = m_3 = m_e$, the
electron mass, and $p_2 = p_3 \simeq p_1/2$. It satisfies the following
equation
\begin{equation}
E_{th} \omega \simeq m_e^2 + \varepsilon \frac{2^a - 1}{2^{2 + a}} 
\frac{E_{th}^{2 + a}}{E_P^a}.
\label{tetev}
\end{equation}
For the general case, the threshold energy $E_{th}$ is given by\cite{ACP}
\begin{equation}
4 E_{th} \omega \simeq (m_2 + m_3)^2 - m_1^2 + \varepsilon 
\frac{E_{th}^{2 + a}}{E_P^a} \left( 1 - \frac{m_2^{1 + a} + m_3^{1 + a}}
{(m_2 + m_3)^{1 + a}}\right),
\label{tegen}
\end{equation}
with $m_1 = m_2 = m_p$, the proton mass and $m_3 = m_{\pi}$, the pion
mass for UHECR.
(One can easily check that Eq. (\ref{tegen}) contains Eq. (\ref{tetev}) as
a special
case.)

To explain the TeV-$\gamma$ events, we need to raise the threshold energy
to $E_{th} \simeq 20 TeV$. With $E_P \simeq 10^{28}$ eV for the Planck 
energy, 
Eq. (\ref{tetev}) gives 
$\varepsilon \simeq 4.2 \times 10^{-5}$ for $a = 2/3$ and $\varepsilon
\simeq 2.5$ for $a = 1$. To explain the UHECR events, we need the threshold shift
from $5 \times 10^{19}$ eV to $E_{th} = 3 \times 10^{20}$ eV; Eq.
(\ref{tegen})
yields $\varepsilon \sim 10^{-17}, 10^{-15}$ for $a = 2/3, 1$ respectively.
Indeed, as expected, $\varepsilon$ is small compared to one in 
general.  (But the smallness of $\varepsilon$ for the UHECR case suggests 
that there may be a fine-tuning problem.  More on the allowed values of 
$\varepsilon$ later.)  It is
amazing that such a small modification coming from energy-momentum
uncertainties can have such a large effect in shifting the threshold
energies by a factor of 2 and 6 for the TeV-$\gamma$ and UHECR events
respectively. To repeat, energy-momentum uncertainties from quantum
gravity effects can potentially be the physical origin of the two threshold anomalies.

The following comment is now in order. Effects of energy-momentum uncertainties 
yield a negative $\varepsilon$ as likely as a positive $\varepsilon$. Then
what happens to the negative $\varepsilon$ case? The answer is that
negative values of $\varepsilon$ would shift the energy thresholds in the
opposite ("wrong") direction. They correspond to events not
seen; therefore, there is nothing that needs to be explained in the first place.

(ii) Modifying the energy-momentum dispersion relation

Consider energy-momentum conservation given by Eq.
(\ref{oemc}) but the energy-momentum dispersion relation modified to read
\begin{equation}
(E_i + \delta E_i)^2 = (p_i + \delta p_i)^2 + m_i^2,
\label{mdr1}
\end{equation}
which, for high energy ($E_i \simeq p_i$), becomes
\begin{equation}
E_i \simeq \frac{1}{2} p_i \left( 2 + \frac{m_i^2}{p_i^2} + \varepsilon 
\frac{p_i^a}{E_P^a}\right),
\label{mdr2}
\end{equation}
where $\varepsilon$ is defined by Eq. (\ref{epsd}) as in scheme (i). 
Eq. (\ref{mdr2}) is the starting point of the approach adopted by many of 
the Lorentz invariance violation advocates\cite{LID,ACP}.  Here it is the 
result of energy-momentum uncertainties (due to quantum gravity) in the 
dispersion relation.  Using Eq. (\ref{mdr2}) and Eq. (\ref{oemc}), we 
recover Eq.
(\ref{lmcr}) for the threshold energy 
except for a sign change for $\varepsilon$. But as we have argued above,
the 
sign of $\varepsilon$ is irrelevant. To raise the threshold energy, all we 
need this time is to pick negative values for $\varepsilon$. 
It follows that, as far as the
UHECR events and TeV-$\gamma$ events are concerned, the threshold
anomalies are explained in the same way as in (i). 

In passing we mention that we have used the same $\varepsilon$ parameter
for all different particle species. If we have used different
$\varepsilon$ parameters for different particle species, we will get a
scheme which bears some resemblance to that advocated by Coleman and
Glashow\cite{colgl}.  (Such dependence of $\varepsilon$ on particle species 
is natural if, e.g., $\delta p_i$ and $\delta E_i$ cancel so completely in 
Eq. (\ref{epsd}) that its right hand side is reduced by a factor of 
$m_i^2/p_i^2$. But in that case, the effect from energy-momentum 
uncertainties is so small that we recover the ordinary threshold equation; 
in other words, we will need another way to solve the threshold anomalies.)  

Are the two schemes (i) and (ii) equivalent? No, not entirely.
Consider the modified energy-momentum dispersion relation for photon given
by Eq. (\ref{mdr2})
\begin{equation}
E^2 \simeq c^2k^2 + \varepsilon E^2 \left(\frac{E}{E_P}\right)^a,
\label{gamd}
\end{equation}
where we have restored c in writing $p = ck$.
The speed of (massless) photon 
\begin{equation}
v = \frac{\partial E}{\partial k} \simeq c \left( 1 + \varepsilon \frac{1 + a}{2} 
\frac{E^a}{E_P^a}\right),
\label{gams}
\end{equation}
becomes energy-dependent! Thus modified energy-momentum dispersion relation (scheme (ii)),
unlike modified energy-momentum conservation relations (scheme (i)), predicts
time-of-flight differences between simultaneously-emitted photons of
different energies, $E_1$ and $E_2$, given by
\begin{equation}
\delta t \simeq \varepsilon t \frac{1 + a}{2} \frac{E_1^a - E_2^a}{E_P^a},
\label{tdiff}
\end{equation}
where $t$ is the average overall time of travel from the photon source. An
upper bound\cite{Ellis,ACP} on the absolute value of $\varepsilon$ can be
obtained
from the observation\cite{tevob} of simultaneous (within experimental
uncertainty of $\delta t \leq 200$ sec) arrival of 1-TeV and 2-TeV
$\gamma$-rays from Mk 421 which is believed to be $\sim 143$ Mpc away from
the Earth. Using Eq. (\ref{tdiff}) we obtain $|\varepsilon| \leq 
1.3 \times
10^{-3}, 1.4 \times 10^2$ for $a = 2/3, 1$ respectively.  Note that these 
bounds for $\varepsilon$ are consistent with those values from UHECR and 
TeV-$\gamma$ events.  For an analysis of the time-lag signature see Ref.\cite{tlsign}.

\section{Discussion}

In the preceding section, we have obtained the various values of 
$\varepsilon$ corresponding to the two observed threshold energies.  But 
an examination of Eqs. (\ref{tetev}) and (\ref{tegen}) shows that, with 
those values of $\varepsilon$, the two equations can each be solved by 
\emph{two} different real and positive $E_{th}$'s the larger of which being 
the observed threshold energy.  Now the following question arises: given 
$\varepsilon$, which of the two solutions for $E_{th}$ would nature pick?  
Perhaps neither.  The point is that, for 
real and positive $\varepsilon$ and $E_{th}$, there is a maximum value of 
$\varepsilon$ above which there is \emph{no} solution to the two equations.  
In that case, the threshold cutoffs are completely removed (i.e., the 
threshold anomalies are trivially solved).  This consideration leads us to 
the following bounds on the (magnitude of the) $\varepsilon$ parameter: 
$\varepsilon \gsim 4.6 \times 10^{-5}, 3.8 \times 10^{-17}$ for the 
TeV-$\gamma$ and UHECR respectively for the case of $a = 2/3$, and 
$\varepsilon \gsim 3.0, 1.5 \times 10^{-14}$ respectively for the case 
$a = 1$. These values of $\varepsilon$ are still consistent with the bounds 
from photon time-of-flight delay measurements given above.

So far we have considered the effects from either modified energy-momentum
conservation relations or a modified energy-momentum dispersion relation.
Let us now consider scheme (iii), the case with both the conservation
relations and the dispersion relation modified. As for scheme (ii),
time-of-flight differences between simultaneously-emitted photons of
different energies are predicted. But as far as the threshold anomalies
are concerned, one can check that this scheme offers no explanation as
the effects of energy-momentum
uncertainties cancel out in the threshold equation, yielding
\begin{equation}
E_{th} = \frac{(m_2 + m_3)^2 - m_1^2}{4 \omega},
\label{othc}
\end{equation}
the ordinary threshold condition which we try to explain away for the UHECR
events and the TeV-$\gamma$ events. (It is not surprising that one gets 
back the ordinary threshold condition for this case because one can
redefine $E_i$ and $p_i$ by absorbing $\delta E_i$ and $\delta p_i$ so that
all energy-momentum uncertainty effects disappear from the threshold equation.)

So, which of the three schemes is the correct one?  Frankly we cannot decide.  However, the three schemes give different experimental
predictions (or "post-dictions"). 
So in principle they can be subject to further experimental
checks.  (But at least we may have provided advocates of Lorentz invariance
violation\cite{LID,colgl,ACP,Ellis} some physical justification coming from
energy-momentum uncertainties due to quantum gravitational fluctuations.)  Our attitude is
that we should proceed in such a way as to preserve as much as possible the 
framework which has been so productive in describing the various
physical interactions.  Thus we would like, on the large scale of
experimental equipment, to preserve time translation-, space translation-,
and Lorentz- invariance.  This would support the familiar conservation laws
to a sufficient extent.  But it does not necessarily mean that space-time in the small
is Minkowskian.  Following Einstein and Wigner we could presumably 
blame small scale oscillations of the metric (which, as we have argued
in Sec. II, lead to energy-momentum uncertainties) for 
possible deformations of
Minkowskian invariance.  Once this fact is accepted, we would expect some
effects in the energy-momentum dispersion relation for the individual
particles participating in a collision as well as in the conservation laws
of energy and linear momentum in such a collision.  At the very least,
we should not accept strict Lorentz
invariance and energy-momentum conservation on faith
but rather regard them as plausible hypotheses subject to experimental 
tests!  Nature may have kindly provided us with the UHECR and TeV-$\gamma$ 
puzzles for such tests.

\bigskip

\begin{center}
{\bf Acknowledgments}\\
\end{center}

One of us (YJN) thanks G. Amelino-Camelia for a useful correspondence and John Ellis for calling to his attention some relevant references. This work 
was 
supported in part by the US Department of Energy under \#DE-FG05-85ER-40219, 
the Bahnson Fund at the 
University of North Carolina at Chapel Hill, and the Republic of China National 
Science Council under NSC89-2112-M-259-001 and NSC89-2112-M-259-012-Y.


\bigskip

\begin{references}

\bibitem{crob}
M.A. Lawrence et al., J. Phys. {\bf G17} (1991) 733; N. N. Efimov et al.,
in 22nd Intl. Cosmic Pay Conf. (1991) Dublin; D.J. Bird et al., Astrophys.
J. {\bf 441} (1995) 144; M. Takeda et al., Phys. Rev. Lett. {\bf 81} (1998)
1163; A. Watson, in Proc. Snowmass Workshop (1996) 126.

\bibitem{GZK}
K. Greisen, Phys. Rev. Lett. {\bf 16} (1966) 748; G. T. Zatsepin and V.A.
Kuz'min, JETP Lett. {\bf 41} (1966) 78.

\bibitem{Oli}
A.V. Olinto, Phys. Rept. {\bf 333-334} (2000) 329-348.

\bibitem{topde}
P. Bhattacharjee, C.T. Hill, and D.N. Schramm, Phys. Rev. Lett. {\bf 69}
(1992) 567; G. Sigl, astro-ph/9611190; V. Berezinsky and A. Vilenkin, Phys.
Rev. Lett. {\bf 79} (1997) 5202.

\bibitem{mono}
T. W. Kephart and T.J. Weiler, Astropart. Phys. {\bf 4} (1996) 271; I.F.M.
Albuquerque, G. R. Farrar, and E.W. Kolb, hep-ph/9805288.

\bibitem{S0}
G.R. Farrar and P.L. Biermann, Phys. Rev. Lett. {\bf 81} (1998) 3579.

\bibitem{Zbur}
D. Fargion, B. Mele, and A. Salis, Ap. J. {\bf 517} (1999) 725; T.J.
Weiler, Astropart. Phys. {\bf 11} (1999) 303; J.L. Crooks, J.O. Dunn, and
P.H. Frampton, astro-ph/0002089.

\bibitem{Relic}
P.H. Frampton, B. Keszthelyi, and Y. J. Ng, Intl. J. Mod. Phys. {\bf D8}
(1999) 117.

\bibitem{gambur}
E. Waxman, Phys. Rev. Lett. {\bf 75} (1995) 386; M. Vietri, Astrophys. J.
{\bf 453} (1995) 883.

\bibitem{spinbh}
E. Boldt and P. Ghosh, Mon. Not. R. Astron. Soc. {\bf 307} (1999) 491.

\bibitem{colga}
C.J. Cesarsky, Nucl. Phys. (Proc. Suppl.) {\bf B28} (1992) 51.

\bibitem{HEGRA}
F.A. Aharonian et al., Astronomy and Astrophysics {\bf 349} (1999) 11A.

\bibitem{PM}
R. J. Protheroe and H. Meyer, astro-ph/0005349, to appear in Phys. Lett. B.

\bibitem{gaexp}
M. Harwit, P.J. Protheroe, and P.L. Biermann, Ap. J. {\bf 524} (1999) L91.

\bibitem{LID}
L. Gonzalez-Mestres, physics/9704017; R. Aloisio, P. Blasi, P.L. Ghia, and
A.F. Grillo, astro-ph/0001258; O. Bertolami and C.S. Carvalho, Phys. Rev.
{\bf D61} (2000) 103002; H. Sato, astro-ph/0005218; T. Kifune, Astrophys.
J. Lett. {\bf 518} (1999) L21; W. Kluzniak, astro-ph/9905308.

\bibitem{colgl}
S. Coleman and S.L. Glashow, Phys. Rev. {\bf D59} (1999) 116008.

\bibitem{ACP}
G. Amelino-Camelia and T. Piran, hep-ph/0006210; astro-ph/0008107.

\bibitem{ng}
Y. J. Ng, gr-qc/0006105.

\bibitem{nvD1}
Y. J. Ng and H. van Dam, Mod. Phys. Lett. {\bf A9} (1994) 335; ibid. {\bf
A10} (1995) 2801; in Proc. of Fundamental Problems in Quantum Theory, eds.
D.M. Greenberger and A. Zeilinger, Ann. New York Acad. Sci. {\bf 755}
(1995) 579.

\bibitem{Ford}
For lightcone fluctuations, see L. H. Ford, Phys. Rev. {\bf D51} (1995) 1692.

\bibitem{MTW}
C. W. Misner, K.S. Thorne, and J.A. Wheeler, Gravitation (W.H. Freeman,
1973), pp. 1190-1194.

\bibitem{KS}
F. Karolyhazy, Il Nuovo Cimento {\bf A42} (1966) 390; N. Sasakura, Prog. Theor. Phys. {\bf 102} (1999) 169, and hep-th/0001161.

\bibitem{nvD2}
Y.J. Ng and H. van Dam, Found. Phys. {\bf 30} (2000) 795; Phys. Lett. {\bf B477} (2000) 429.

\bibitem{IJMP}
G. Amelino-Camelia, J. Ellis, N.E. Mavromatos, and D.V. Nanopoulos, Int. J. Mod. Phys. {\bf A12} (1997) 607.

\bibitem{Ellis}
G. Amelino-Camelia, J. Ellis, N.E. Mavromatos, D.V. Nanopoulos, and S.
Sarkar, Nature {\bf 393} (1998) 763.  The idea of bounding $\varepsilon$ originated with this paper.

\bibitem{tevob}
B. E. Schaefer, Phys. Rev. Lett. {\bf 82} (1999) 4964; S.D. Biller et al.,
Phys. Rev. Lett. {\bf 83} (1999) 2108.

\bibitem{tlsign}
J. Ellis, K. Farakos, N.E. Mavromatos, V.A. Mitsou, and D.V. Nanopoulos, Astrophys. J. {\bf 535} (2000) 139.

\end{references}
\end{document}